\documentclass[11pt,a4paper]{article}
\pdfoutput=1   % required for arXiv submission if pdflatex is used
\usepackage[utf8]{inputenc}
\usepackage{tabularx}
\usepackage{mathtools}
\usepackage{longtable}
\usepackage{alpha} 
\hypersetup{%
   pdftitle    = {Mass-improvement of the vector current in three-flavor QCD},
   pdfauthor   = {Patrick Fritzsch},
   pdfkeywords = {Lattice QCD, Nonperturbative Effects, Symanzik improvement}
}%
\usepackage[outercaption]{sidecap}
\usepackage{defs}
\begin{document}

\preprintno{%
CERN-TH-2018-119\\
\vfill
}

\title{%
Mass-improvement of the vector current \\
in three-flavor QCD 
}

\author[cern]{Patrick Fritzsch}
\address[cern]{Theoretical Physics Department, CERN, 1211 Geneva 23, Switzerland}

\begin{abstract}
We determine two improvement coefficients which are relevant to cancel
mass-dependent cutoff effects in correlation functions with operator insertions
of the non-singlet local QCD vector current.  This determination is based on
degenerate three-flavor QCD simulations of non-perturbatively O($a$) improved
Wilson fermions with tree-level improved gauge action.  Employing a very robust
strategy that has been pioneered in the quenched approximation leads to an
accurate estimate of a counterterm cancelling dynamical quark cutoff effects
linear in the trace of the quark mass matrix.  To our knowledge this is the
first time that such an effect has been determined systematically with large
significance.\\
\end{abstract}

\begin{keyword}
Lattice QCD \sep Nonperturbative Effects \sep Symanzik improvement %
\PACS{% 
11.15.Ha\sep %Lattice Gauge Theory
11.30.Rd\sep %Chirality - particle physics
12.38.Gc\sep %Lattice QCD calculations
12.38.Aw     %general properties of QCD
}
\end{keyword}

\maketitle

\tableofcontents
%\clearpage

%
%         File : s1.tex
%      Created : Wed 28 Mar 2018 10:09:57 AM CEST
%  Last Change : Fri 18 May 2018 04:37:55 PM CEST
%
\section{Prelude}\label{sec:intro}

Lattice QCD has reached a maturity in which some quantities can be computed
very precisely such that QED or isospin breaking effects become relevant when
compared to experiment. Other quantities are being improved in various ways
to match the accuracy reached in state-of-the art experiments, or to use it
as input in the search for new physics.

In either case, the control over continuum extrapolations of lattice data is an
important ingredient that contributes to the overall precision in any
observable. If the QCD action is discretized using Wilson
fermions~\cite{Wilson:1974sk}, chiral symmetry is broken explicitly by a term
proportional to the lattice spacing $a$, and consequently (renormalized)
observables receive contributions linear in the lattice spacing. With some
additional effort these terms can be systematically removed as outlined by
Symanzik~\cite{Symanzik:1983dc,Symanzik:1983gh,Symanzik:1981hc}. In the present
paper we aim at a computation of the coefficients $\bV$ and $\bbV$ which
multiply mass-dependent counterterms to restore a continuum scaling to
$\rmO(a^2)$ in the local vector current made of massive quarks. Thus they are
relevant in calculations such as semileptonic vector form factors or the muon
anomalous magnetic moment.

In cases where the impact of an improvement coefficient seems negligible from a
perturbative point of view, it can still be important to have a
non-perturbative estimate at hand in order to properly judge at what level of
accuracy they will become important. To determine such coefficients one
requires well-justified improvement conditions. These conditions can be chosen
freely to some extent but differ in complexity and quality regarding
statistical and/or systematic effects---like ambiguities at higher order in
$a$.  They all remove the leading scaling violation they have been designed
for, sometimes at the cost of introducing large contributions at higher orders.
The condition explored in the present paper relies on the QCD isospin vector
symmetry, and we argue that this method, adopted from
reference~\cite{Luscher:1996jn}, has little systematic effects. Being
implemented in the QCD Schr\"odinger functional
(SF)~\cite{Luscher:1992an,Sint:1993un,Luscher:1984xn,Aoki:1998qd} it
furthermore allows to efficiently explore the region about the chiral limit
over a wide range of couplings and masses at relatively small computational
costs.  As a result we are able to differentiate between the valence and
dynamical (sea) quark sector when the gauge coupling becomes strong.

The pattern of chiral symmetry breaking, inherent to Wilson fermions at finite
lattice spacing, intertwines renormalization and $\rmO(a)$
improvement~\cite{Luscher:1996sc}. As a result, the local (isovector) vector
current, $V_{\mu}^a(x)=\psibar(x) \gamma_{\mu} \tfrac{1}{2}\tau^a \psi(x)$,
of mass-degenerate fermions renormalizes as%
\footnote{Although we are interested in $\Nf>2$ quark flavors, the relevant
          subgroup for quark bilinears is $SU(2)$.
}%
\begin{align}\label{eq:VR}
        (V_{\rm\subX R})_\mu^a &= \ZV(\gtilde^2) \left( 1 + \bV(g_0^2) a\mq  + \bbV(g_0^2) \Tr[a\Mq] \right) (V_{\rm\subX I})_\mu^a  \;, \\
        (V_{\rm\subX I})_\mu^a &= V_\mu^a + \cV(g_0^2) a\tilde\partial_\mu T_{\mu\nu}^a \;, \notag
\end{align}
in a \emph{massless renormalization scheme}~\cite{Luscher:1996jn}. Here,
$V_{\rm\subX I}$ is the bare vector current, on-shell $\rmO(a)$ improved through
the symmetric lattice derivative of the tensor current $T_{\mu\nu}^a$ with
appropriate improvement coefficient $\cV$. We follow the standard notation in
the literature~\cite{Luscher:1996sc,Bhattacharya:2005rb} where $c_{*}$ terms
cancel $\rmO(a)$ cutoff effects and $b_{*}$, $\bar{b}_{*}$ the leading
mass-dependent effects of order $am$, induced from the valence or sea quark
sector, respectively.%
\footnote{In massive schemes it can be convenient to absorb all mass effects
into $Z$-factors and $c$-coeff., cf.~\cite{Fritzsch:2018kjg}.
}
These coefficients are considered functions of the bare gauge coupling $g_0^2$
while the finite renormalization $\ZV$ has to be evaluated at $\gtilde^2=g_0^2(
1 + \bg(g_0^2)\Tr[a\Mq]/\Nf)$~\cite{Luscher:1996sc,Sint:1995ch,Sint:1997jx} to
maintain full $\rmO(a)$ improvement of $V_{\rm\subX R}$ in the presence of
massive quarks. Rather being a mass-dependent finite renormalization than a
genuine lattice artefact, the coupling counterterm coefficient $\bg(g_0^2)$ is
required to cancel the mass-dependence of $g_0^2$ in such a way that
$\gtilde^2$ is mass-independent up to terms of order $a^2$. Note that both
couplings coincide at the critical line.
From eq.~\eqref{eq:VR} it then becomes clear that $\bV$ and $\bbV$ can be
determined by individual varying the valence quark mass $a\mq$ and sea quark
mass $\Tr[a\Mq]$.

%
%         File : s2.tex
%      Created : Wed 28 Mar 2018 10:15:25 AM CEST
%  Last Change : Wed 06 Jun 2018 09:02:23 AM CEST
%
%-----------------------------------------------------------------------------
\section{Centerpiece}\label{sec:strategy}
%-----------------------------------------------------------------------------

We are simulating $\Nf=3$ mass-degenerate flavors of Wilson fermions with
tree-level improved gauge action in the Schr\"odinger functional, exhibiting
Dirichlet boundary conditions in time and periodic boundary conditions in
spatial directions.  For on-shell $\rmO(a)$ improvement of the action we use
the non-perturbatively determined clover coefficient
$\csw$~\cite{Sheikholeslami:1985ij} of ref.~\cite{Bulava:2013cta} and one-loop
values of the boundary counterterms $\ct$ and
$\cttil$~\cite{Luscher:1996sc,Takeda:2003he,DallaBrida:2016kgh}.  Except for
simulating at non-vanishing quark mass, our setup thus equals the one of
ref.~\cite{DallaBrida:2016kgh} with vanishing boundary fields, $T=L$ and
$\theta=1/2$. In the following we adopt the notation of
refs.~\cite{Luscher:1996sc,Luscher:1996jn,Hoffmann:2005cz}.

%-----------------------------------------------------------------------------
\subsection{Improvement condition of the vector current}

For completeness we first introduce the matrix elements used in the present
calculation and comment on possible systematic effects. In the SF sources are
typically defined on the time boundaries at Euclidean times $x_0=0$ and
$x_0=T$, 
\begin{align}\label{eq:bSRC}
        \clO^a          &= a^6 \sum_{\vecu,\vecv} \zetabar(\vecu)        \gamma_{5} \tfrac{1}{2}\tau^a \zeta(\vecv) \;, & 
        \clO^{\prime a} &= a^6 \sum_{\vecu,\vecv} \zetabar^\prime(\vecu) \gamma_{5} \tfrac{1}{2}\tau^a \zeta^\prime(\vecv)  \;,
\end{align}
respectively. They are projected to zero momentum and transform according to
the vector representation of the exact isospin symmetry.
The simplest correlation function is the two-point boundary-to-boundary
correlator
\begin{align}\label{eq:fone}
        \fone &= -\frac{1}{3L^6} \langle  \clO^{\prime a}\clO^a \rangle
\end{align}
which up to terms of order $a^2$ equals the three-point boundary-to-boundary
correlator
\begin{align}\label{eq:fVR}
        f_{\mathrm{V_{R}}} &= \frac{a^3}{6L^6} \sum_{\vecx} i\epsilon_{abc} \langle  \clO^{\prime a} (V_{\rm\subX R})_0^b \clO^c \rangle
\end{align}
with an insertion of the time component of the renormalized vector current,
eq.~\eqref{eq:VR}. It is important to notice that (a) the normalization and
improvement of the sources are irrelevant and (b) the $\rmO(a)$ counterterm
proportional to $\cV$ cancels as well. Thus two potential sources to
$\rmO(a^2)$ ambiguities are absent and the condition becomes
\begin{align}\label{eq:cond}
   \fone &= \ZV \left( 1 + \bV a\mq  + \bbV \Tr[a\Mq] \right) \fV(x_0) + \rmO(a^2)
\end{align}
where $\fV$ as in eq.~\eqref{eq:fVR} after $(V_{\rm\subX R})_{0}^b\to V_{0}^b$
and $\ZV$ the overall normalization in the chiral limit.  It also implies that
(c) the relation is independent of $x_0$ up to cutoff effects, or more
specifically, independent of $\ct$ and $\cttil$. However, if not stated
otherwise we take $x_0=T/2$ as the preferred definition of
relation~\eqref{eq:cond}.  Our restricted interest in the mass-improvement
coefficients leads to another simplification.  Estimators of $\bV$ and $\bbV$
can be realised as derivatives such that (d) the overall normalization factor
$\ZV$ cancels, e.g.,
\begin{align}\label{eq:RV}
        \RV &= \frac{\partial}{\partial a\mq} \left.\log\left( \frac{\fone}{\fV} \right)\right|_{\mq=0}  \;, & \text{at}~\Mq &= 0{\text{ and } g_0^2~\text{fixed}}.
\end{align}
The derivative essentially eliminates all mass-independent cutoff effects such
that $\RV= \bV + \rmO(am)$ mainly carries ambiguities that depend on the
quark mass.%
\footnote{If the term proportional to $\cV$ would be relevant, it would
          contribute at $\rmO(am)$ to $\RV$.
}
If the derivative is computed in a range of quark masses where $\fone/\fV$ is
dominantly linear, these ambiguities are practically absent as (e) it does not
matter at which point the derivative is exactly evaluated. 
As a consequence of (d)+(e) we conclude that one does not necessarily need to
impose a line of constant physics when determining $\RV$. 

Concerning sea quark mass effects we remark that in the unitary case
($\mq=\msea$) with $\Nf$ degenerate quarks ($[\Mq]_{ij}=\delta_{ij}\msea$ with
$i,j=1,\ldots,\Nf$) only the combination $\bV+\Nf \bbV$ is accessible from
eq.~\eqref{eq:cond}. We thus define
\begin{align}\label{eq:barRV}
        \barRV &= \frac{\partial}{\partial a\msea} \left.\log\left( \frac{\fone}{\fV} \right)\right|_{\msea=\mq}\!\!  \;, & \text{at}~\Mq &= 0 {\text{ and } g_0^2~\text{fixed}},
\end{align}
as estimator for $\bV + \Nf \bbV$ and obtain $\bbV$ trivially from an
appropriate linear combination with $\RV$.
This being said, we proceed as follows: 
(1) at different values of the lattice spacing we use a given set of dynamical
three-flavor simulations with varying sea quark mass to determine the chiral
limit ($\msea=0$) by simple interpolation, 
(2) we determine the combination $\bV+\Nf\bbV$ for the unitary case via
eq.~\eqref{eq:barRV},
(3) on the ensemble in the vicinity of vanishing quark mass we vary the valence
quark masses in order to determine the mass-derivative~\eqref{eq:RV} and thus
$\bV$. With above arguments about the potential quality of the improvement
conditions~(\ref{eq:RV},\ref{eq:barRV}), and to keep the numerical effort
and costs affordable, we restrict our simulations to lattices of size $L/a=8$.
We show explicitly that this is sufficient in the present case.  Our data
analysis accounts for autocorrelations via the
$\Gamma$-method~\cite{Wolff:2003sm} and correlations are included through
standard linear error propagation.

%-----------------------------------------------------------------------------
\subsection{Chiral trajectory and current quark mass in finite volume}\label{s2:chiral}
\begin{SCtable}[][!t]
    \small
    \centering
    \begin{tabular}{crllll}\toprule
    $L/a$ & $\beta$ & $\hopcr$       & $\bar{Z}$    \\\midrule 
      $8$ & $32.0$  & $0.126209(1) $ & $1.020(1) $  \\  
      $8$ & $16.0$  & $0.127496(1) $ & $1.033(2) $  \\  
      $8$ & $ 8.0$  & $0.130387(2) $ & $1.073(5) $  \\  
      $8$ & $ 4.0$  & $0.136648(3) $ & $1.226(26)$  \\  
      $8$ & $ 3.7$  & $0.137053(3) $ & $1.283(34)$  \\  
      $8$ & $ 3.6$  & $0.137059(5) $ & $1.318(43)$  \\  
      $8$ & $ 3.5$  & $0.136949(4) $ & $1.414(14)$  \\  
      $8$ & $ 3.4$  & $0.136669(4) $ & $1.566(69)$  \\  
      $8$ & $ 3.3$  & $0.136160(11)$ & $2.02(13) $  \\  
      \bottomrule
    \end{tabular}
    \caption{\label{tab:kcr_sea}%
             Values of the critical hopping parameters $\hopcr$ and slopes
             $\barZ$ at the unitary point, obtained by linearly interpolating
             data in the range $|Lm|<0.22$, see table~\ref{tab:Msea}. The
             resulting uncertainty is combined linearly with the absolute
             difference to the corresponding values obtained by removing the
             outermost data points from the least square minimization.
            }   
\end{SCtable}

As usual we define the chiral limit as the point where the current quark
mass $m$ vanishes.%
\footnote{We define $m$ as in section A.2.1 of~\cite{Capitani:1998mq} with
          non-perturbative $\cA$ of ref.~\cite{Bulava:2015bxa}.
}
For simplicity we restrict the discussion to the unitary
case ($\mq=\msea$) and remark that overall (multiplicative) renormalization
factors are not required. In the $\rmO(a)$ improved theory, the relation
between current quark mass $m$ and bare subtracted sea quark mass $\msea$ reads
\begin{align}\label{eq:m_msea}
        m &= \bar{Z} \msea(1+B\, a\msea) + \rmO(a^2)  \;,
\end{align}
where $\bar{Z}=Z\Zrm$ is a relative (finite) normalization and $B$ a
well-defined combination of mass-improvement
coefficients~\cite{Bhattacharya:2005rb}. With similar numerical simulations
close to the chiral limit at hand~\cite{Nf3tuning}, we do apply one iteration
of refinements at 9 different values of the gauge coupling $\beta=6/g_0^2$ to
perform simulations in close proximity of $m=0$. In all cases we are as close
as $|am|<0.001$ and aim for a symmetric variation in the quark mass spanning
several orders of magnitude to probe the regime of linearity of
eq.~\eqref{eq:m_msea} with significant precision. Where it is possible we go as
far as $Lm\approx\pm 2$, but only compile the subset most relevant for the
determination of $\RV$ and $\barRV$ in table~\ref{tab:Msea}, where
$a\msea=\tfrac{1}{2}(\hopsea^{-1}-\hopcr^{-1})$ is the bare subtracted quark
mass based on the critical hopping parameters $\hopcr$ of
table~\ref{tab:kcr_sea}. By steadily increasing the number of points about
vanishing current quark mass entering a least square minimisation to a straight
line in $1/(2\hopsea)$, we inspect the quality of the fit to decide on the
range of linearity.  In all cases, we do not find relevant deviations in the
range $|a\msea|< 0.014 $. As this is of minor importance anyway, we determine
$\hopcr$ by fitting to data in that range only.  For completeness we give the
corresponding slope, $\bar{Z}$, as estimate for the potential size of $Z\Zrm$,
but have to remind the reader that this quantity is better determined
along a line of constant physics.  We present a typical interpolation in the
top panel of figure~\ref{fig:beta}, compiled from data of table~\ref{tab:Msea}
at an intermediate coupling ($\beta=4.0$) and strong coupling ($\beta=3.4$).
For the latter we know the size of the lattice spacing, $a\approx 0.086\,\fm$,
from ref.~\cite{Bruno:2016plf}. The bare subtracted quark mass at $a\msea=0.1$
thus corresponds to $\msea\approx 230\,\MeV$, and at $a\msea=0.01$ accordingly
to $23\,\MeV$. To state the obvious: within the given uncertainty it is
confirmed that the improvement $B$-term in eq.~\eqref{eq:m_msea} is negligible
for up-/down-like quarks while it can be relevant for strange-like quarks, and
certainly is for heavier ones. 

%-----------------------------------------------------------------------------
\subsection{Unitary quark mass dependence}\label{s2:res_unitary}

\begin{SCfigure}[][!t]
    \small
    \includegraphics[width=0.6\textwidth]{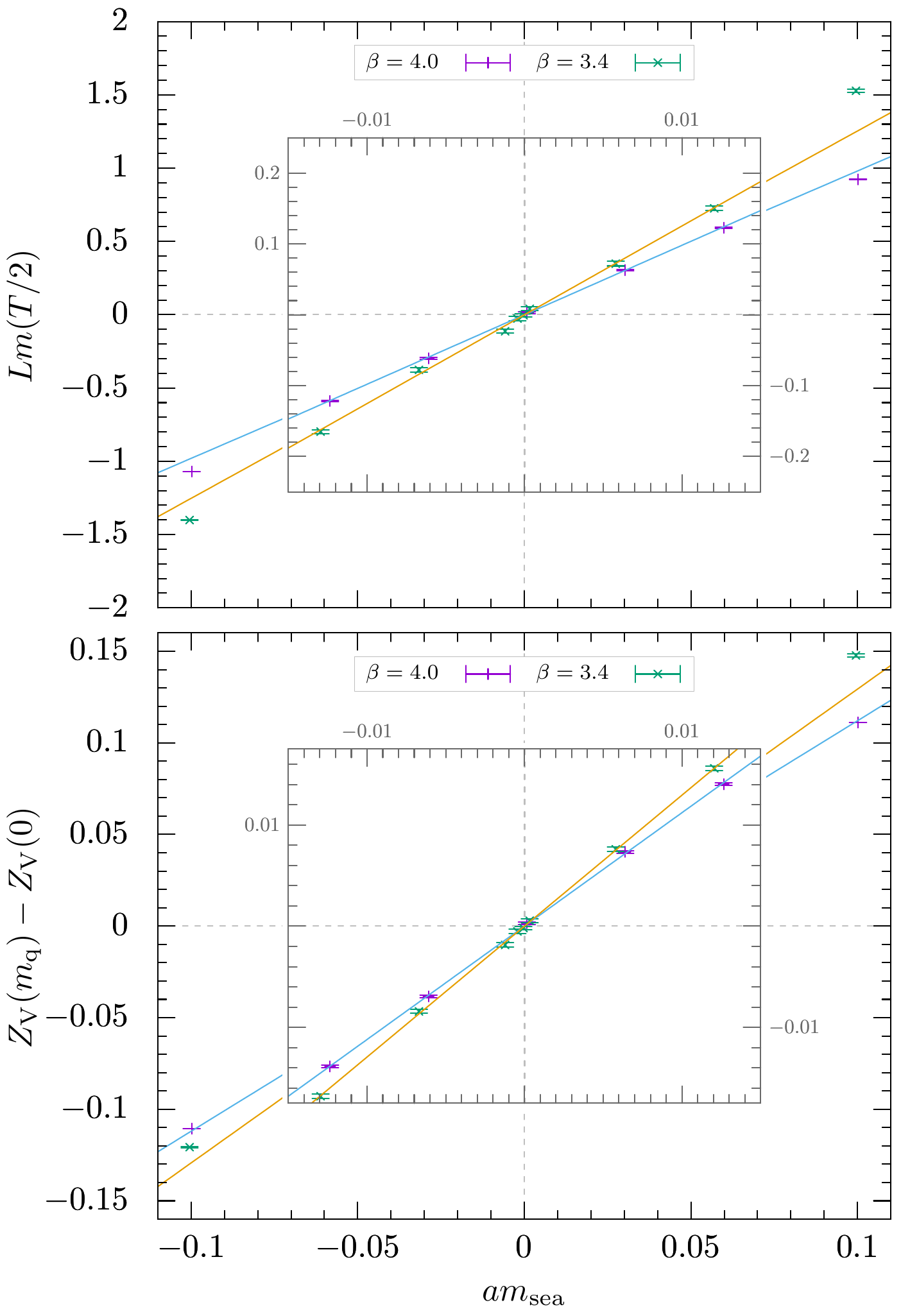}
    \caption{\label{fig:beta}%
             Sea quark mass-depen\-dence ($\mq=\msea$) of the current quark
             mass $Lm$ and renormalization factor $\ZV$ with typical
             interpolating functions at $\beta=4.0, 3.4$. The inner graph
             is a magnification of data entering the straight line fit
             close to the chiral limit ($|a\mq|<0.014$), while the outer
             graph visualises the deviation from linear behaviour far away
             from the chiral limit. All data points are independent from
             each other and taken from table~\ref{tab:Msea}.  The quality
             of fit tends to be somewhat better for $\ZV$ data when
             compared to $Lm$ data. Note that the fit quality of $Lm$ at
             $\beta=3.4$ improves when the points atop $|a\mq|=0.012$ are
             included. 
            }
\end{SCfigure}

We have measured the correlation functions $\fone$ and $\fV$ in the unitary
setup for all sea quark masses. Their values are listed in table~\ref{tab:Msea}
in dependence of $\msea$ and we use $\ZV(\mq)$ as short hand for the
mass-dependent ratio $\fone(\mq)/\fV(\mq)$.  We clearly profit from
correlations between the two boundary-to-boundary correlation functions but
note that values on different ensembles are uncorrelated. As in the previous
section we study the range in which the data is well compatible with a straight
line fit ansatz, which again is the case for $|a\msea|< 0.014 $, cf. bottom
panel of figure~\ref{fig:beta}.  At each coupling $\beta$ we neglect the
ensemble closest to the chiral limit in order to be fully independent from the
determination of $\RV$ presented in the next section.  The associated
interpolation provides the intercept at $\msea=0$ and mass-derivative
$\partial_{\mq}\ZV(\mq)$. Combined they give estimates of $\barRV$ as listed in
table~\ref{tab:barRV}. Its uncertainty is about one percent and below. For
reasons described in the next section, we increase the uncertainty of $\barRV$
at $\beta=3.3$ before probing various trial Pad\'e functions in a weighted
least square minimisation.  The data is best described by 
\begin{align}\label{eq:barRV_fit}
      \barRV(g_0^2) &= \frac{1+\bar{p}_{1}g_0^2+\bar{p}_{2}g_0^4+\bar{p}_{3}g_0^6}{1+\bar{p}_{4}g_0^2}  \;, &
      (\bar{p}_{i}) &= \begin{pmatrix}  -0.43101  \\ 
                                        +0.04109  \\ 
                                        -0.03911  \\
                                        -0.51771
                       \end{pmatrix} \;,
\end{align}
which we take as our preferred representation of $\bV+\Nf\bbV$. With the 
accuracy achieved in the present paper, its uncertainty
is not really of any practical relevance, except when we derive the 
uncertainty of $\bbV$ in section~\ref{s2:res_sea}. For completeness we 
quote ($i,j=1,\ldots,4$)
\begin{align}\label{eq:barRV_cov}
  \cov(\bar{p}_{i},\bar{p}_{j}) &=   
         \begin{pmatrix*}[r] 4.9676 & -4.8982 &  3.0473  & 3.2513   \\  
                             *      &  6.4514 & -3.6515  & -2.6826  \\  
                             *      &  *      &  2.1830  &  1.8785  \\
                             *      &  *      & *        &  2.4558  \\
         \end{pmatrix*} \times 10^{-5} \;.
\end{align}
The functional form~\eqref{eq:barRV_fit} is plotted in
figure~\ref{fig:interpol} together with the input data. We should remark that
the improvement coefficient $\bV$ is known to one-loop order in perturbation
theory~\cite{Aoki:1998qd,Taniguchi:1998pf}, $\bVpt(g_0^2)=1+0.0886(2)\CF
g_0^2=1+0.1181(3) g_0^2$.  As the leading contribution of $\bbV$ is at
$\rmO(g_0^4)$, we could constrain all Pad\'e ans\"atze by $\bVpt(g_0^2)$.
However, the accuracy of our data, especially at small couplings ($\beta=32.0,
16.0$), reveals some tension to the one-loop estimate which is hard to account
for in that case: $\bar{p}_1-\bar{p}_4=0.087(3)$. This could be due to the 
asymptotic nature of the perturbative series when compared to a non-perturbative
result, or may expose a point we have not stressed explicitly yet. We have
argued that the normalization conditions in use have hardly any systematic
effects. On the other hand they explicitly depend on the non-perturbative
clover-coefficient $\csw$~\cite{Bulava:2013cta} that renders the Wilson action
$\rmO(a)$ improved.  Although the Pad\'e ansatz used for $\csw(g_0^2)$ accounts
for the correct one-loop coefficient, it undershoots it at couplings below
$g_0^2\approx 0.9$ and approaches it only asymptotically. In that respect it is
similar to what we observe but without data in the deep perturbative regime.

Of course, this would have gone unnoticed as any fit excluding our $\beta=32.0,
16.0$ data can be easily constrained by the aforementioned one-loop $\bVpt$. In
the end, this discussion is of purely academic interest as no simulation of
physical interest is ever performed in that region of couplings. But it is the
reason why we do not insist in constraining our final result,
eq.~\eqref{eq:barRV_fit}.

%-----------------------------------------------------------------------------
\subsection{Valence quark mass dependence}\label{s2:res_valence}

Next we come to our determination of $\bV$ which is required in a partially
quenched setup, e.g., with charmed valence quarks. For that purpose, we choose
the ensemble closest to vanishing quark mass at each coupling, and for
convenience measure $\fone$ and $\fV$ with hopping parameters $\hopval$
coinciding with the $\hopsea$ of other simulations listed in
table~\ref{tab:Msea}. As a result, the measurements of $\ZV(\mq)$ for different
valence quark masses are strongly correlated, leading to much more accurate
estimates of $\RV$. Even with higher accuracy, we do not observe mentionable
differences from linearity in the range $|a\mq|<0.014$ and as in the previous
section stick to it in our final analysis. Before continuing we want to present
a stringent test to explicitly validate the use of $L/a=8$ lattices. 

\begin{SCfigure}[][!tp]
        \small
        \centering
        \includegraphics[width=0.6\textwidth]{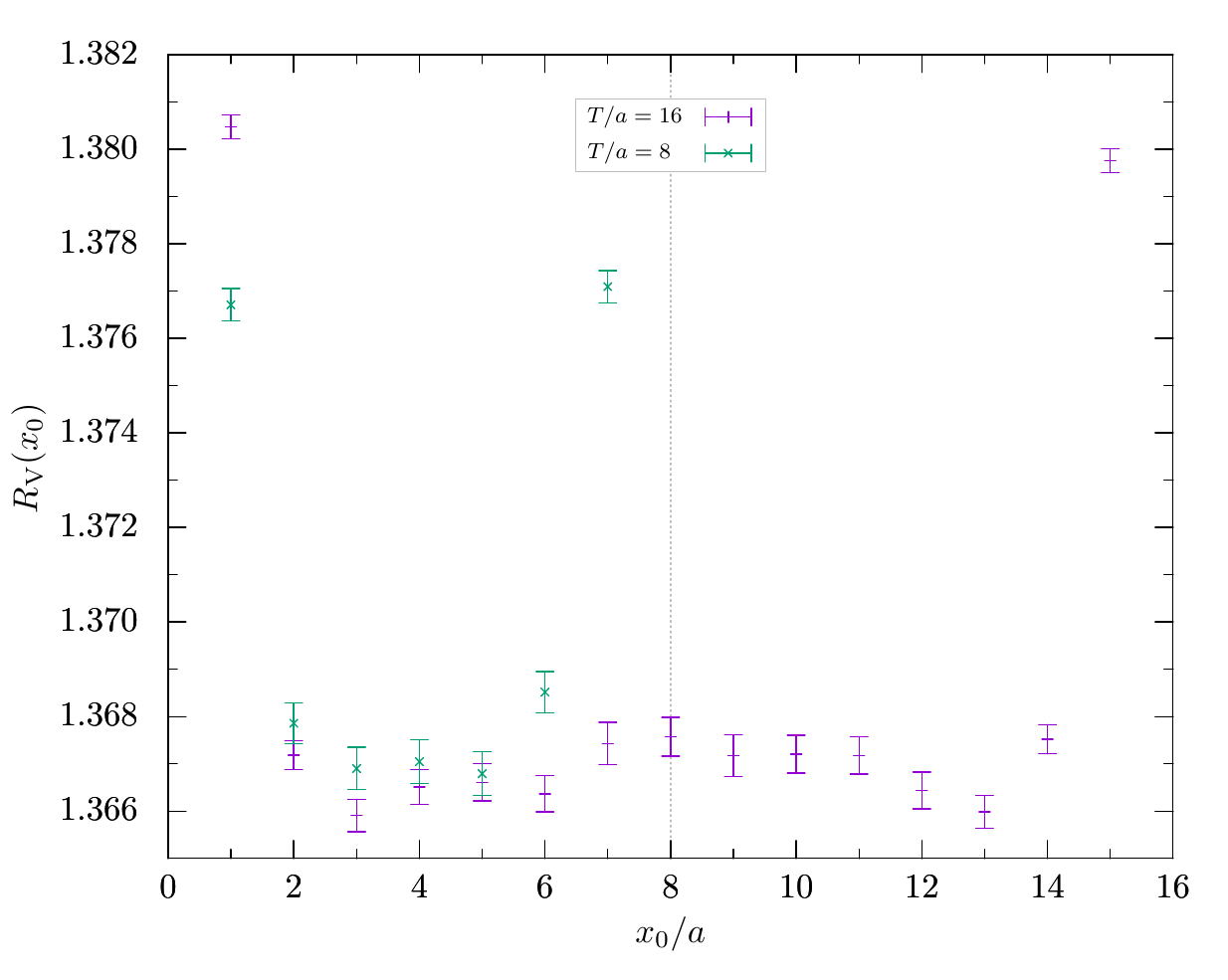}
        \caption{Euclidean time (in)de\-pendence of $\RV$ at $\beta=4.0$ for
                 $L/a=8, 16$ lattices.  Straight line interpolations take all
                 correlations into account for the data shown here. Except for
                 the points closest to the boundaries, there is no
                 statistically significant $x_0$-dependence.
                }
        \label{fig:beta4}
\end{SCfigure}

In figure~\ref{fig:beta4} we show the Euclidean time dependence of $\RV$ at
$\beta=4.0$ where two volumes are at our disposal, $T=L=8a,16a$. The latter is
available to us thanks to the work in
ref.~\cite{DallaBrida:2016kgh,Campos:2018ahf}, and we refer to it for further
details about that simulation. First, it confirms that, one or two lattice
units away from the boundaries, there is no $x_0$-dependence beyond statistical
fluctuations, suggesting that $\rmO(a^2)$ effects in eq.~\eqref{eq:cond}
are small. Second, the agreement of $\RV(T/2)$ between both volumes shows that
our estimate is volume independent. These observations hold within the
statistical uncertainty which is at the per mille level or below.  Monitoring
the $x_0$-dependence for all values of $\beta$, we remark that only at the
coarsest lattice spacing ($\beta=3.3$) we observe a $+2\sigma$ deviation in
both $\RV(T/2+a)$ and $\RV(T/2-a)$. To account for this still insignificant
fluctuation, we conservatively double the uncertainty of $\RV(T/2)$.

We compile results of $\RV$ for two types of analysis in table~\ref{tab:barRV}.
For reasons that become clear soon, we distinguish between a full (correlated)
analysis and one neglecting correlations between measurements at different
masses, which increases the uncertainty of $\RV$ by about an order of
magnitude.  Of course, both are in full agreement. In the course of simulations
to produce ensembles at different sea quark masses we were guided by the
statistical precision in $\barRV$, leading to a precision in the correlated
analysis of $\RV$ that goes far beyond what is needed in future applications.
While it allowed to probe relation~\eqref{eq:cond} and the range of linearity
more precisely than before, it is not straightforward to find a proper
global approximation at that level of precision without increasing the number
of data points.
\begin{figure}[t]
   \small
   \centering
   \includegraphics[width=0.8\textwidth]{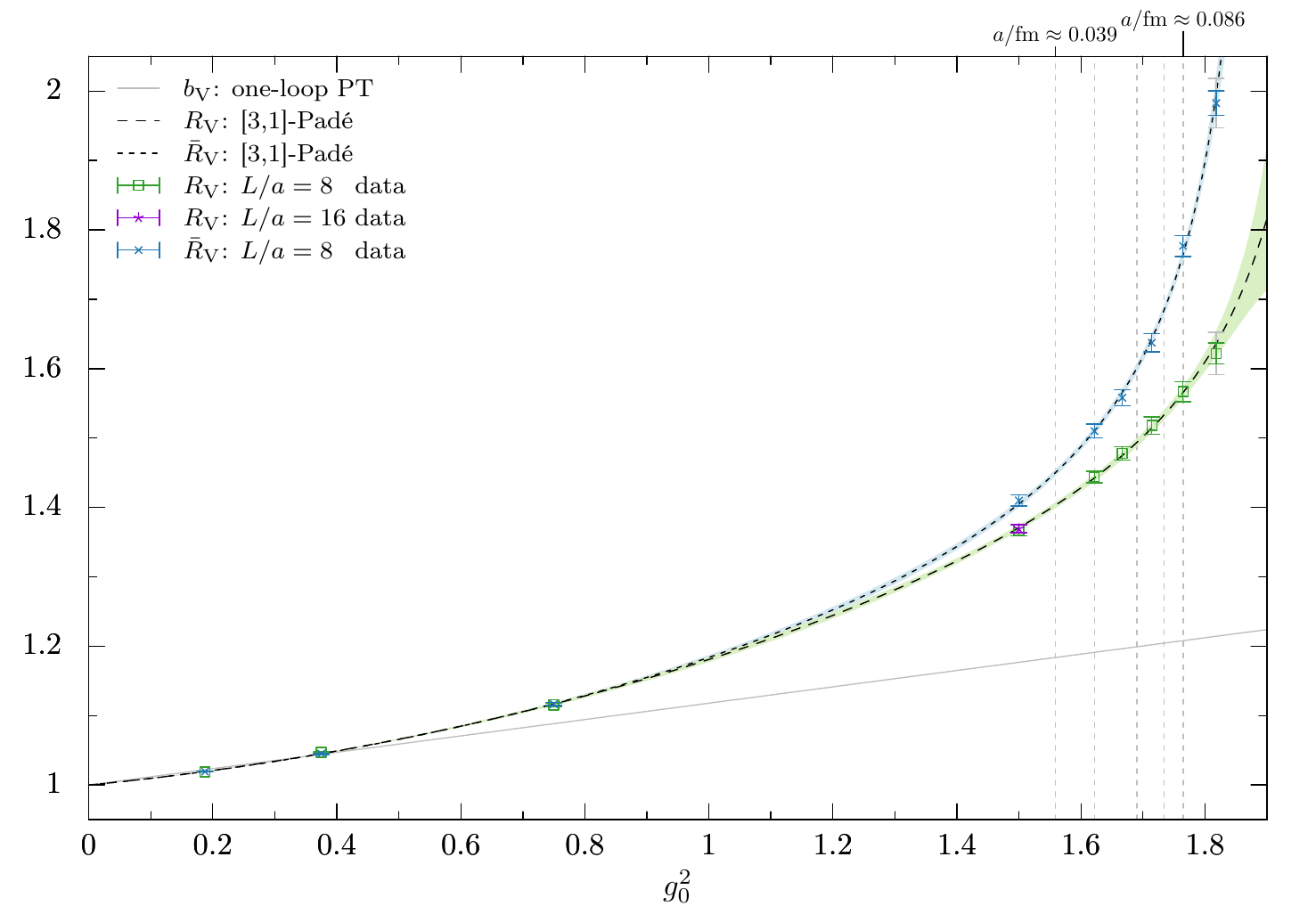}
   \caption{\label{fig:interpol}%
            Data with interpolating function for $\RV$ and $\barRV$ in
            comparison to the known perturbative one-loop estimate $\bVpt$,
            see text. Vertical dashed lines mark values of
            $\beta\in\{3.4,3.46,3.55,3.7,3.85\}$ used in large volume
            simulations~\cite{Bruno:2014jqa,Bruno:2016plf,Bruno:2017gxd},
            with indicated largest and smallest lattice spacing $a$.
           }
\end{figure}

For the sake of simplicity we thus prefer to employ data from the uncorrelated
analysis on each ensemble, remarking once more that results at different
couplings are statistically independent. At $\beta=4.0$ we furthermore include 
the independent data point from $L/a=16$.  Not surprisingly, a $[3,1]$-Pad\'e
function gives the best result that reads
\begin{align}\label{eq:RV_fit}
        \RV(g_0^2) &= \frac{1+p_1g_0^2+p_2g_0^4+p_3g_0^6}{1+p_4g_0^2} \;, &
         ({p}_{i}) &= \begin{pmatrix}  -0.40040  \\ 
                                       +0.04352  \\ 
                                       -0.03852  \\
                                       -0.48803  
                      \end{pmatrix}  \;,
\end{align}
with parameter covariance matrix
\begin{align}\label{eq:RV_cov}
  \cov({p}_{i},{p}_{j}) &=   
         \begin{pmatrix*}[r] 86.1424  &  -20.5770  &  23.7151  &  78.9559  \\  
                             *        &    8.1121  &  -6.9724  & -17.7160  \\  
                             *        &  *         &   7.1199  &  21.3673  \\
                             *        &  *         & *         &  72.9602  \\
         \end{pmatrix*}  \times 10^{-5}  \;.
\end{align}
We add this interpolation and the entering data points to
figure~\ref{fig:interpol}. One observes a significant difference to $\barRV$ in
the strong coupling region where (2+1)-flavor large volume simulations with the
same bulk lattice action, produced by the Coordinated Lattice Simulations
effort (CLS), exist~\cite{Bruno:2014jqa,Bruno:2016plf,Mohler:2017wnb}.  Based
on those simulations various estimates of mass-improvement coefficients have
been published in ref.~\cite{Korcyl:2016ugy} for
$\beta\in\{3.4,3.46,3.55,3.7\}$.  The uncertainty of the latter for $\bV$ is of
the size of our estimated difference between sea and valence quark sector.

%-----------------------------------------------------------------------------
\subsection{On sea quark mass-effects in the vector channel}\label{s2:res_sea}

\begin{figure}[t]
     \small
     \centering
     \includegraphics[width=0.8\textwidth]{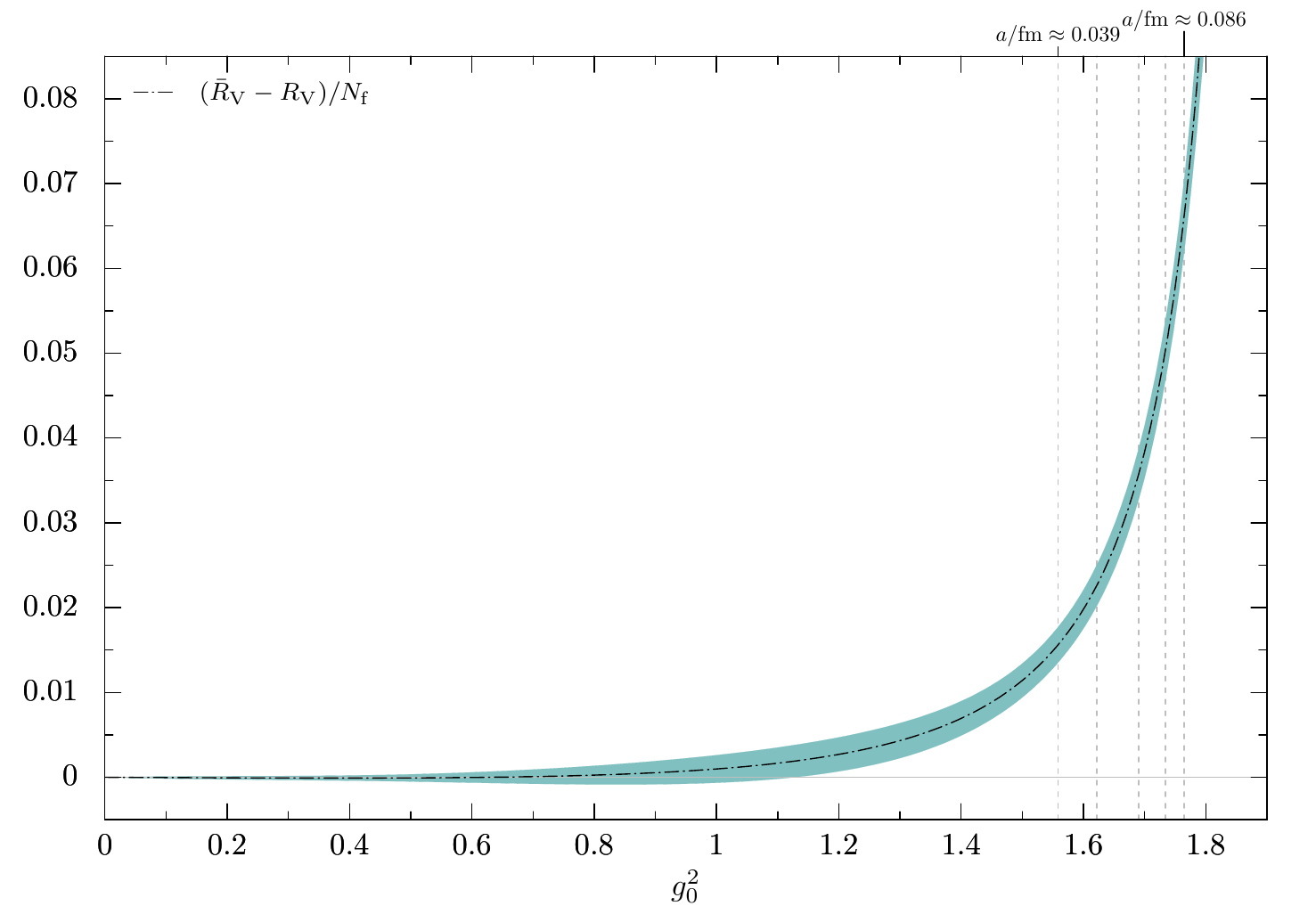}
     \caption{\label{fig:barbV}%
              Sea quark mass-improvement coefficient $\bbV=(\barRV-\RV)/\Nf$.
             }
\end{figure}

With interpolating formulae for $\RV$ and $\barRV$ we are finally able to check
on the size of pure sea quark mass-effects that are cancelled at leading order
by the counterterm $\bbV\Tr[a\Mq]$. We define
\begin{align}\label{eq:bV}
   \bbV(g^2_0) &= \left( \barRV(g_0^2)-\RV(g_0^2) \right)/\Nf
\end{align}
by eq.~(\ref{eq:barRV_fit},\ref{eq:RV_fit}) and present it in
figure~\ref{fig:barbV}.
$\barRV$ and $\RV$ are statistically independent, such that the uncertainty
of $\bbV$ follows trivially from standard Gaussian error propagation.
The value of $\bbV$ is different from zero in the non-perturbative region of
large volume simulations~\cite{Bruno:2014jqa}, being about $0.07$ at
$\beta=3.4$. For those simulations the unsubtracted trace of the quark mass
matrix has been kept constant w.r.t. the $SU(3)$-flavor symmetric point such
that the physics at fixed coupling $g_0^2$ and $\Tr[a\Mq]$, and thus fixed
$\gtilde^2$, does not require any knowledge of $\bg$.  With $\sum_{i=\rm
u,d,s}(2\kappa_i)^{-1}\approx 10.97$ at $\beta=3.4$ and an preliminary estimate
$\hopcr(\beta)\approx 0.13698$ of ref.~\cite{Nf3tuning} this implies
$\Tr[a\Mq]\approx 0.02$. The resulting counterterm to cancel mass-dependent sea
quark cutoff effects of a strange quark thus becomes an effect at the per mille
level. For a dynamical heavy charm quark in the sea this effect would rise to
the per cent level even without including higher order effects.  

Given the similarity between $\bA$, $\bP$ and $\bV$, cf.~\cite{Sint:1997jx},
one could be lead to the assumption that $\bbA$ and $\bbP$ are comparable in
size to the result for $\bbV$ presented above, but only a direct computation
can confirm that.

%
%         File : s3.tex
%      Created : Tue 08 May 2018 09:39:28 AM CEST
%  Last Change : Fri 18 May 2018 12:33:49 PM CEST
%
\section{Epilogue}

In combination with a precise determination of the overall renormalization
factor $\ZV$ and improvement coefficient $\cV$ along a line of constant
physics~\cite{Brida:2014zwa,Heitger:2017njs,Nf3xSF:ZA}, the determination of
$\bbV$ and $\bV$ presented in this paper completes the on-shell $\rmO(a)$
improvement of the non-singlet local vector current with massive Wilson
fermions that is consistent with massless renormalization schemes. Our results
are independent from determinations of $\ZV$ and $\cV$, and thanks to the
applied improvement conditions we have reached very accurate estimates of $\barRV$
and $\RV$ that allowed for disentangling sea and valence quark effects
with large significance for the first time. These estimates are particularly
relevant to lattice phenomenology on CLS (2+1)-flavor ensembles with lattice
spacings in the range $0.04\lesssim a/\fm\lesssim0.09$.
Due to the smallness of $\bbV$ even at the non-perturbative level ($a\lesssim
0.1\,\fm$)  it seems likely that these effects are negligible in most
applications of lattice QCD with dynamical light quarks. We have shown
explicitly that higher order mass-dependent effects are practically absent, 
which is not true anymore for, e.g., dynamical charm quarks.

\begin{acknowledgement}%
We thank M.~L\"uscher for a critical reading of the manuscript.  P.F.
acknowledges financial support by the Spanish MINECO’s “Centro de Excelencia
Severo Ochoa” Programme under grant SEV-2012-0249, as well as from the MCINN
grant FPA2012-31686. The main part of this work has profited from the IFT
computing infrastructure, especially the Hydra cluster, and some additional
measurements were performed on the CERN cluster.  We thank both IT departments
for the provided resources and support.
\end{acknowledgement}

\begin{appendix}
%
%         File : a1.tex
%      Created : Tue 15 May 2018 10:17:33 AM CEST
%  Last Change : Tue 15 May 2018 11:41:09 AM CEST
%
\addcontentsline{toc}{section}{Miscellaneous}
\section*{Miscellaneous}

Let there be table~\ref{tab:Msea} and~\ref{tab:barRV}.
\vskip1em

{
\small
\begin{longtable}{lllllll}
       \caption{\label{tab:Msea}%
                Unitary simulation points with corresponding current quark mass
                and correlator results. After thermalization, the majority of
                simulations has accummulated ensemble sizes of 5000
                configurations or more, separated by 4 molecular dynamic units
                each.  Measurements have been performed on each configuration.
               }\\[-0.5em]
       \toprule
       $\beta$ & $\hopsea$    & $a\msea$  &  $Lm$            &  $\fone(\mq)$  & $\fV(\mq)|_{x_0=T/2}$ &  $\ZV(\mq)$ \\\midrule
           32  & $0.12581342$ & $+0.0125$ &  $+0.101367(54)$ &  $1.23284(22)$ &  $1.23354(22)$   &  $0.999436(9)$   \\ 
               & $0.12601159$ & $+0.0062$ &  $+0.050765(52)$ &  $1.32366(28)$ &  $1.33279(29)$   &  $0.993145(9)$   \\ 
               & $0.12621039$ & $-0.0000$ &  $+0.000059(53)$ &  $1.42134(29)$ &  $1.44025(30)$   &  $0.986870(9)$   \\ 
               & $0.12640982$ & $-0.0063$ &  $-0.051145(55)$ &  $1.52593(33)$ &  $1.55616(34)$   &  $0.980575(9)$   \\ 
               & $0.12660988$ & $-0.0125$ &  $-0.102681(55)$ &  $1.63694(33)$ &  $1.68015(35)$   &  $0.974285(9)$   \\\midrule
           16  & $0.12709186$ & $+0.0125$ &  $+0.10271(12)$  &  $1.11557(45)$ &  $1.13923(47)$   &  $0.979236(18)$  \\ 
               & $0.12729408$ & $+0.0062$ &  $+0.05158(13)$  &  $1.20192(49)$ &  $1.23538(51)$   &  $0.972914(17)$  \\ 
               & $0.12749695$ & $-0.0000$ &  $+0.00006(12)$  &  $1.29321(56)$ &  $1.33789(60)$   &  $0.966599(20)$  \\ 
               & $0.12770047$ & $-0.0063$ &  $-0.05186(12)$  &  $1.39299(60)$ &  $1.45057(64)$   &  $0.960303(18)$  \\ 
               & $0.12790464$ & $-0.0125$ &  $-0.10388(12)$  &  $1.49863(69)$ &  $1.57094(75)$   &  $0.953969(19)$  \\\midrule 
           8   & $0.12996592$ & $+0.0124$ &  $+0.10633(29)$  &  $0.89988(96)$ &  $0.9641(10)$    &  $0.933404(42)$  \\   
               & $0.13017740$ & $+0.0062$ &  $+0.05294(30)$  &  $0.9772(10)$  &  $1.0541(11)$    &  $0.926998(43)$  \\
               & $0.13038957$ & $-0.0001$ &  $+0.00002(29)$  &  $1.0566(11)$  &  $1.1477(12)$    &  $0.920627(40)$  \\
               & $0.13060244$ & $-0.0063$ &  $-0.05432(36)$  &  $1.1459(15)$  &  $1.2535(16)$    &  $0.914155(53)$  \\
               & $0.13060244$ & $-0.0063$ &  $-0.05348(45)$  &  $1.1477(19)$  &  $1.2555(21)$    &  $0.914138(58)$  \\
               & $0.13081600$ & $-0.0126$ &  $-0.10846(29)$  &  $1.2468(14)$  &  $1.3735(16)$    &  $0.907710(40)$  \\\midrule
           4   & $0.13300823$ & $+0.1007$ &  $+0.9229(11)$   &  $0.1059(4)$   &  $0.1169(4)$     &  $0.90578(12)$   \\
               & $0.13617797$ & $+0.0132$ &  $+0.1228(11)$   &  $0.5150(21)$  &  $0.6369(26)$    &  $0.80868(12)$   \\
               & $0.13641017$ & $+0.0070$ &  $+0.0631(11)$   &  $0.5735(25)$  &  $0.7151(32)$    &  $0.80201(13)$   \\ 
               & $0.13664316$ & $+0.0007$ &  $+0.0028(11)$   &  $0.6462(28)$  &  $0.8129(35)$    &  $0.79497(12)$   \\ 
               & $0.13687696$ & $-0.0055$ &  $-0.0615(12)$   &  $0.7339(32)$  &  $0.9316(42)$    &  $0.78773(13)$   \\ 
               & $0.13711155$ & $-0.0118$ &  $-0.1216(12)$   &  $0.8106(37)$  &  $1.0381(48)$    &  $0.78080(12)$   \\
               & $0.14048236$ & $-0.0993$ &  $-1.0699(14)$   &  $3.270(37)$   &  $4.781(55)$     &  $0.68408(14)$   \\\midrule
           3.7 & $0.13659066$ & $+0.0124$ &  $+0.1264(15)$   &  $0.4593(21)$  &  $0.5876(27)$    &  $0.78168(14)$   \\
               & $0.13682427$ & $+0.0061$ &  $+0.0625(14)$   &  $0.5245(25)$  &  $0.6774(33)$    &  $0.77429(15)$   \\ 
               & $0.13702933$ & $+0.0007$ &  $+0.0063(16)$   &  $0.5929(34)$  &  $0.7721(45)$    &  $0.76794(16)$   \\ 
               & $0.13705868$ & $-0.0001$ &  $+0.0002(15)$   &  $0.5984(35)$  &  $0.7800(46)$    &  $0.76723(17)$   \\ 
               & $0.13705868$ & $-0.0001$ &  $-0.0021(13)$   &  $0.5983(24)$  &  $0.7799(32)$    &  $0.76719(12)$   \\ 
               & $0.13708804$ & $-0.0009$ &  $-0.0096(15)$   &  $0.6156(32)$  &  $0.8035(43)$    &  $0.76606(15)$   \\ 
               & $0.13729390$ & $-0.0064$ &  $-0.0632(15)$   &  $0.6815(42)$  &  $0.8966(56)$    &  $0.76015(15)$   \\ 
               & $0.13752992$ & $-0.0126$ &  $-0.1318(16)$   &  $0.7791(46)$  &  $1.0353(62)$    &  $0.75254(17)$   \\\midrule 
           3.6 & $0.13659365$ & $+0.0125$ &  $+0.1303(18)$   &  $0.4466(26)$  &  $0.5797(35)$    &  $0.77043(17)$   \\
               & $0.13682727$ & $+0.0062$ &  $+0.0650(18)$   &  $0.5149(29)$  &  $0.6746(39)$    &  $0.76326(16)$   \\ 
               & $0.13703235$ & $+0.0008$ &  $+0.0075(19)$   &  $0.5844(35)$  &  $0.7723(47)$    &  $0.75668(17)$   \\ 
               & $0.13706169$ & $-0.0000$ &  $-0.0011(15)$   &  $0.5876(28)$  &  $0.7775(37)$    &  $0.75574(14)$   \\ 
               & $0.13709105$ & $-0.0008$ &  $-0.0076(18)$   &  $0.6004(36)$  &  $0.7950(48)$    &  $0.75514(18)$   \\ 
               & $0.13729692$ & $-0.0063$ &  $-0.0639(18)$   &  $0.6781(42)$  &  $0.9060(56)$    &  $0.74848(16)$   \\
               & $0.13753296$ & $-0.0125$ &  $-0.1348(17)$   &  $0.7799(48)$  &  $1.0525(66)$    &  $0.74100(17)$   \\\midrule 
           3.5 & $0.13647158$ & $+0.0127$ &  $+0.1448(21)$   &  $0.4202(28)$  &  $0.5539(37)$    &  $0.75864(18)$   \\ 
               & $0.13670478$ & $+0.0065$ &  $+0.0726(24)$   &  $0.4980(37)$  &  $0.6631(50)$    &  $0.75107(21)$   \\ 
               & $0.13693878$ & $+0.0002$ &  $+0.0038(22)$   &  $0.5721(41)$  &  $0.7698(55)$    &  $0.74310(17)$   \\ 
               & $0.13717359$ & $-0.0060$ &  $-0.0687(23)$   &  $0.6770(47)$  &  $0.9203(65)$    &  $0.73562(20)$   \\ 
               & $0.13740920$ & $-0.0123$ &  $-0.1380(22)$   &  $0.7883(59)$  &  $1.0823(82)$    &  $0.72836(20)$   \\\midrule 
           3.4 & $0.13304903$ & $+0.0996$ &  $+1.527(11)$    &  $0.0077(4)$   &  $0.0088(4)$     &  $0.87560(87)$   \\
               & $0.13622073$ & $+0.0121$ &  $+0.1502(33)$   &  $0.4129(43)$  &  $0.5554(59)$    &  $0.74346(22)$   \\
               & $0.13645308$ & $+0.0058$ &  $+0.0721(32)$   &  $0.4990(44)$  &  $0.6785(60)$    &  $0.73544(21)$   \\ 
               & $0.13665703$ & $+0.0003$ &  $+0.0085(29)$   &  $0.5804(50)$  &  $0.7969(69)$    &  $0.72839(19)$   \\ 
               & $0.13667180$ & $-0.0001$ &  $-0.0005(24)$   &  $0.5896(35)$  &  $0.8102(49)$    &  $0.72767(16)$   \\ 
               & $0.13668622$ & $-0.0004$ &  $-0.0052(30)$   &  $0.5934(53)$  &  $0.8158(74)$    &  $0.72734(22)$   \\ 
               & $0.13671542$ & $-0.0012$ &  $-0.0233(30)$   &  $0.6138(50)$  &  $0.8454(70)$    &  $0.72601(21)$   \\ 
               & $0.13692016$ & $-0.0067$ &  $-0.0777(31)$   &  $0.7066(55)$  &  $0.9822(78)$    &  $0.71944(21)$   \\
               & $0.13715490$ & $-0.0129$ &  $-0.1649(28)$   &  $0.8364(74)$  &  $1.176(10)$     &  $0.71105(22)$   \\
               & $0.14052786$ & $-0.1004$ &  $-1.4014(30)$   &  $3.12(60)$    &  $5.139(99)$     &  $0.60712(29)$   \\\midrule 
           3.3 & $0.13573478$ & $+0.0114$ &  $+0.1933(65)$   &  $0.3815(71)$  &  $0.525(10)$     &  $0.72595(24)$   \\
               & $0.13596547$ & $+0.0051$ &  $+0.0870(58)$   &  $0.5073(72)$  &  $0.708(10)$     &  $0.71697(23)$   \\ 
               & $0.13613900$ & $+0.0005$ &  $+0.0068(46)$   &  $0.6076(63)$  &  $0.8558(89)$    &  $0.70995(22)$   \\ 
               & $0.13619695$ & $-0.0011$ &  $-0.0225(42)$   &  $0.6539(64)$  &  $0.9242(92)$    &  $0.70761(23)$   \\ 
               & $0.13625494$ & $-0.0027$ &  $-0.0444(41)$   &  $0.6875(75)$  &  $0.974(11)$     &  $0.70555(21)$   \\ 
               & $0.13642921$ & $-0.0074$ &  $-0.1052(52)$   &  $0.7913(98)$  &  $1.131(14)$     &  $0.69959(27)$   \\
               & $0.13666227$ & $-0.0136$ &  $-0.2181(47)$   &  $0.988(11)$   &  $1.431(15)$     &  $0.69083(27)$   \\\midrule 
\end{longtable}
}

\begin{table}[!h]
        \small
        \centering
        \caption{\label{tab:barRV}%
                Summary of results and data analysis for $\barRV$ and $\RV$,
                using input data in the range $|a\mq|<0.014$. Note that the
                simulation of $L/a=16$ at $\beta=4.0$ has $\kappa=0.13668396$
                and $Lm=0.0006(8)$, cf. table~10 of ref.~\cite{Campos:2018ahf}.
                (top rows: uncorrelated analysis, bottom rows:
                fully correlated analysis)\\[-1.5em]
                }
        \begin{tabular}{CLLLLLLL}\toprule
            &       & \multicolumn{3}{c}{\text{sea sector}}     & \multicolumn{3}{c}{\text{valence sector}} \\\cmidrule(lr){3-5}\cmidrule(lr){6-8}
       L/a  & \beta & \ZV(0)       & \partial_m\ZV & \barRV     & \ZV(0)       &\partial_m\ZV& \RV          \\\midrule
         8  &  32   & 0.986902(4)  & 1.0059(5)     & 1.0193(5)  & 0.986909(4)  & 1.0057(5)   & 1.0190(5)    \\ 
         8  &  16   & 0.966639(8)  & 1.0103(10)    & 1.0452(10) & 0.966633(9)  & 1.0119(10)  & 1.0469(10)   \\ 
         8  &  8    & 0.920652(18) & 1.0279(21)    & 1.1165(22) & 0.920647(24) & 1.0269(27)  & 1.1154(29)   \\ 
         8  &  4    & 0.794704(56) & 1.1205(63)    & 1.4099(79) & 0.794802(54) & 1.0865(62)  & 1.3670(77)   \\ 
         16 &  4    & -            & -             &  -         & 0.786608(20) & 1.0772(46)  & 1.3694(58)   \\
         8  &  3.7  & 0.767303(54) & 1.1586(78)    & 1.510(10)  & 0.767307(47) & 1.1078(63)  & 1.4437(82)   \\ 
         8  &  3.6  & 0.755889(62) & 1.1777(85)    & 1.558(11)  & 0.755754(53) & 1.1168(71)  & 1.4777(94)   \\ 
         8  &  3.5  & 0.743022(85) & 1.2165(97)    & 1.637(13)  & 0.742886(83) & 1.1277(94)  & 1.518(13)    \\ 
         8  &  3.4  & 0.727888(72) & 1.293(11)     & 1.777(15)  & 0.727381(73) & 1.140(10)   & 1.567(14)    \\ 
         8  &  3.3  & 0.709415(89) & 1.406(13)     & 1.983(18)  & 0.708629(81) & 1.149(11)   & 1.622(15)    \\\midrule 
         8  &  32   & -            & -             & -          & 0.98691(1)   & 1.00566(2)  & 1.01900(2)   \\ 
         8  &  16   & -            & -             & -          & 0.96663(2)   & 1.01195(3)  & 1.04688(4)   \\ 
         8  &  8    & -            & -             & -          & 0.92065(5)   & 1.02686(9)  & 1.1154(1)    \\ 
         8  &  4    & -            & -             & -          & 0.79480(12)  & 1.0865(3)   & 1.3670(5)    \\ 
         16 &  4    & -            & -             & -          & 0.78762(5)   & 1.0771(3)   & 1.3676(4)    \\ 
         8  &  3.7  & -            & -             & -          & 0.76731(13)  & 1.1078(4)   & 1.4437(6)    \\ 
         8  &  3.6  & -            & -             & -          & 0.75575(14)  & 1.1167(5)   & 1.4777(7)    \\ 
         8  &  3.5  & -            & -             & -          & 0.74289(18)  & 1.1277(7)   & 1.5181(10)   \\ 
         8  &  3.4  & -            & -             & -          & 0.72738(20)  & 1.1405(8)   & 1.5679(14)   \\ 
         8  &  3.3  & -            & -             & -          & 0.70864(21)  & 1.1516(9)   & 1.6251(15)   \\ 
         \bottomrule
        \end{tabular}
\end{table}

\end{appendix}

\small
\addcontentsline{toc}{section}{References}
\bibliographystyle{JHEP_erratum}
\bibliography{mainbib}

\end{document}